\documentclass[aps,twocolumn,groupedaddress,showpacs,floatfix]{revtex4}

\usepackage{graphicx}

\begin{document}
\bibliographystyle{apsrev}

\title{A Variational Monte Carlo Study of the Current Carried by a Quasiparticle}

\author{Cody P. Nave$^1$, Dmitri A. Ivanov$^2$ and Patrick A. Lee$^1$}
\affiliation{
$^{1}$Department of Physics, Massachusetts Institute of Technology,
Cambridge, Massachusetts 02139\\
$^{2}$Institute of Theoretical Physics, \'Ecole Polytechnique F\'ed\'erale de Lausanne (EPFL), 
CH-1015 Lausanne, Switzerland}
\date{September 30, 2005}

\begin{abstract}
With the use of Gutzwiller-projected variational states, we study the renormalization
of the current carried by the quasiparticles in high-temperature superconductors 
and of the quasiparticle spectral weight.
The renormalization coefficients are computed by the variational Monte Carlo technique,
under the assumption that quasiparticle excitations may be described by Gutzwiller-projected
BCS quasiparticles. We find that the current renormalization coefficient decreases with
decreasing doping and tends to zero at zero doping. The quasiparticle spectral weight $Z_+$ 
for adding an electron shows an interesting structure in $\mathbf{k}$ space, which 
corresponds to a depression of the occupation number $n_\mathbf{k}$ just outside the Fermi surface.  
The perturbative corrections to those quantities in the Hubbard model are also discussed.
\end{abstract}


\maketitle

\section{Introduction}
In recent years, it has been acknowledged that ground-state properties of high-temperature
superconductors may be reasonably well described with the help of Gutzwiller-projected
wave functions \cite{Anderson:04}. However the main challenge of any candidate theory of high-temperature
superconductivity is the description of finite-temperature properties such as the superconducting
transition and the pseudogap phenomenon in underdoped cuprates. One of the first issues
related to the finite-temperature physics of high-temperature superconductors is the
structure of low-lying excitations. Within the framework of Gutzwiller-projected wave functions,
the first steps in studying the excitations have been recently made: the quasiparticle
spectrum and the quasiparticle spectral weight have been 
calculated \cite{Paramekanti:01,Sorella:05}. In our paper,
we complement the previous studies with the analysis of the current carried by the 
quasiparticles. The magnitude of the quasiparticle current has a direct physical implication
in reducing the superfluid density at finite temperature, which eventually determines
the superconducting transition temperature in the underdoped regime \cite{Lee:97,Wen:98}. Furthermore,
the deviation of the quasiparticle current from the prediction of the BCS theory may provide
a helpful insight in the physics of high-temperature superconductivity.

The reduction of the superfluid density  $n_s(T)/m$  by thermal quasiparticles at the
nodes of a $d$-wave superconductor has been computed by Lee and Wen \cite{Lee:97,Wen:98} as
\begin{equation}
\frac{\hbar^2 n_s(T)}{m} = \frac{\hbar^2 n_s(0)}{m} - \frac{2\ln2}{\pi}\alpha^2
(\frac{v_F}{v_{\Delta}})k_BT \, ,
\label{maineq1}
\end{equation}
where $v_F$ and $v_\Delta$ are the velocity of the nodal quasiparticles 
perpendicular and parallel to the
underlying Fermi surface, and $\alpha$ is the phenomenological Landau parameter \cite{Millis:98} which 
renormalizes the current carried by the quasiparticle
\begin{equation}
\mathbf{j}(\mathbf{k}) = - e \alpha \mathbf{v}_F.
\end{equation}
Experimentally, $n_s(T)/m$ can be related to the London penetration depth $\lambda$, and
the ratio $v_F / v_\Delta$ may be extracted independently from a thermal-conductivity 
measurement \cite{Chiao:00}.
This makes $\alpha$ an experimentally accessible quantity for a variety of 
doping values \cite{Liang:05,Broun:05}.

In the first part of the paper, we focus on computing the current renormalization $\alpha$ for
different doping values. We find that it decreases with decreasing doping, and that it is
roughly constant along the Fermi surface at all dopings. 
The contribution of particles and holes to the total quasiparticle current
allows us to picture the ``effective Fermi surface'' where the 
electron contribution crosses over to the hole contribution. We observe that this
``effective Fermi surface'' deviates considerably from the original Fermi surface
of the unprojected BCS state. This reveals the particle-hole asymmetry produced
by the Gutzwiller projection.

In the second part of the paper, we discuss another renormalization parameter: the
quasiparticle spectral weight $Z_+$ for adding an electron. The momentum dependence
of $Z_+$ shows a pocket structure at the diagonal of the Brillouin zone
just outside the Fermi surface. We further discuss the relations and bounds on
$Z_+$ in the $t$--$J$ model, as well as
corrections arizing from the rotation to the Hubbard model.

For our analysis, we take the minimal two dimensional model for strongly 
interacting electrons on a lattice, the Hubbard model. Following the usual
procedure (see, e.g., Refs.\ \onlinecite{Paramekanti:01,Paramekanti:04}), 
we first study the wave function for its strong-coupling
limit, the $t$--$J$ model and then include the first-order correction in $t/U$
due to the doubly-occupied sites. The Hamiltonian for the $t$--$J$ model
for our system is defined on the two-dimensional square lattice by
\begin{eqnarray}
\mathcal{H}_{tJ} &=& -t\sum_{\langle i,j \rangle,\sigma}\left( c_{i,\sigma}^{\dagger} c_{j,\sigma} +
\mathrm{h.c.} \right) \nonumber\\
&& +J\sum_{\langle i,j \rangle}\left(\mathbf{S}_i \cdot \mathbf{S}_j - n_i n_j/4\right),
\label{tjh}
\end{eqnarray}
with $t/J=3$.  Here $c_{i,\sigma}^{\dagger}$ is the electron creation operator at site $i$ with spin
$\sigma = (\uparrow,\downarrow)$.  The $\langle i,j \rangle$ indicates nearest neighbors, 
$\mathbf{S}_i = \frac{1}{2} c_{i,\alpha}^{\dagger} \vec{\sigma}_{\alpha,\beta} c_{i,\beta}$,
and $n_i= c_{i,\alpha}^{\dagger} c_{i,\alpha}$. 
The Hamiltonian (\ref{tjh}) is then supplemented with the constraint that no double occupancy is allowed.

We use the variational Monte Carlo technique to calculate expectation values of 
operators given our trial wave functions. \cite{Gros:89,Yokoyama:88}
For the $t$--$J$ model, we consider two related trial wave functions:
the ground state wave function $|\Psi_{\mathrm{GS}}\rangle$ and 
the wave function for the excited state $|\Psi_{\mathrm{EX}}\rangle$.  For the ground state,
we use the Gutzwiller-projected $d$-wave singlet, 
\begin{equation}
|\Psi_{\mathrm{GS}}\rangle = P_D P_N | \Psi_{\mathrm{BCS}}\rangle
\end{equation}
where $P_D=\prod_{i} \left[1-n_{i,\uparrow} n_{i,\downarrow}\right]$ is the Gutzwiller 
projection operator onto the subspace with no doubly occupied states, and
$P_N$ is the  projection operator onto the subspace with $N$ particles.
$|\Psi_{\mathrm{BCS}}\rangle = \prod_{\mathbf{k}} \left[ 1 +a_\mathbf{k} c_{\mathbf{k},\uparrow}^\dagger 
c_{-\mathbf{k},\downarrow}^\dagger \right] |0\rangle$, where $c_{\mathbf{k},\sigma}^{\dagger}$ is
the Fourier transform of real-space electron creation operator $c_{i,\sigma}^{\dagger}$.
Following the standard BCS definitions for a $d$-wave singlet state,
\begin{eqnarray}
&a_\mathbf{k}&=\frac{v_\mathbf{k}}{u_\mathbf{k}}=\frac{\Delta_\mathbf{k}}
{\xi_\mathbf{k}+\sqrt{\xi_\mathbf{k}^2+\Delta_\mathbf{k}^2}},\\
&\Delta_\mathbf{k}& = \Delta_{\mathrm{var}}(\cos{k_x}-\cos{k_y}),\\
&\xi_\mathbf{k}&=\varepsilon_\mathbf{k}-\mu_{\mathrm{var}},
\end{eqnarray}
with $\varepsilon_\mathbf{k}=-2(\cos{k_x}+\cos{k_y})$.  
Not only has this $d$-wave Gutzwiller-projected wave function been shown to give good variational 
energies for the $t$--$J$ model relative to other possible phases, 
but also it has correctly reproduced many properties 
of the superconducting state.\cite{Yokoyama:96,Paramekanti:01}

For the trial wavefunction of the low-lying excited states, we take the natural ansatz of
the Gutzwiller-projected Bogoliubov quasiparticle (\cite{Anderson:87,Sorella:05})
\begin{equation}
|\Psi_{\mathrm{EX}}(\mathbf{k},\sigma)\rangle = P_D P_N \gamma_{\mathbf{k},\sigma}^\dagger |\Psi_{\mathrm{BCS}}\rangle.
\label{psiex1}
\end{equation}  
Since the overall normalization of the wave function is of no importance,
we may also rewrite the trial excited state as
\begin{equation}
|\Psi_{\mathrm{EX}}(\mathbf{k},\sigma)\rangle = P_D P_N c_{\mathbf{k},\sigma}^\dagger 
|\Psi_{\mathrm{BCS}}\rangle .
\end{equation}
Throughout the paper, we  suppress the $\mathbf{k}$ and $\sigma$ variables on 
$|\Psi_{\mathrm{EX}}\rangle$ for notational convenience. 
The expectation values of any operator $\mathcal{O}$ in the variational
ground and excited states will be often denoted as
$\langle \mathcal{O}\rangle_{\mathrm{GS}}$ and $\langle \mathcal{O}\rangle_{\mathrm{EX}}$
respectively.

In our simulations, we use the optimal values of $\Delta_\mathrm{var}$ 
and $\mu_\mathrm{var}$ calculated for the ground state trial wave function \cite{Ivanov:03},
both for the ground state and for the excited state. These values minimize the expectation
value $\langle \mathcal{H}_{tJ} \rangle_{\mathrm{GS}}$ of the physical $t$--$J$ Hamiltonian
at a fixed concentration of holes.

We assume boundary conditions that are anti-periodic in the $x$ direction and periodic in the $y$ direction 
so that we avoid the singularity in $a_\mathbf{k}$
along the nodal diagonal, $(0,0)$ to $(\pi,\pi)$.  A drawback of this choice of boundary conditions
is that we are unable to calculate quantities exactly on the nodal diagonal,
and instead calculate expectation values for nearby $\mathbf{k}$ points.  This becomes an important
source of error when we compute expectation values at the nodal point.  We try to 
lower this error by looking at larger systems in order to get better resolution; however, we are 
limited by our computing resources.

This trial excited state and various related ones have been studied previously by other 
groups \cite{Giamarchi:93,TKLee:97,TKLee:03,Sorella:05}.
The energy dispersion of the low-energy quasiparticles has been found to be of the BCS type 
$E(\mathbf{k})= \left[ \xi^2+\Delta_\mathbf{k}^2 \right]^{1/2}$, but with renormalized values 
of the gap and bandwidth. The nodal point obviously coincides with the nodal point of the
unprojected wave function $\Psi_{\mathrm{BCS}}$ and slowly shifts inwards from  $(\pi/2,\pi/2)$
along the diagonal of the Brillouin zone as the hole doping increases.

\section{CURRENT CARRIED BY THE QUASIPARTICLES}
\label{section:current}

In this section, we investigate the current carried by quasiparticles 
as a function of
their momenta and doping.  The current carried by the
excited state $|\Psi_{\mathrm{EX}}(\mathbf{k},\sigma)\rangle$ is defined as
\begin{equation}
\langle \mathbf{j}_{\mathbf{k},\sigma}\rangle = \langle \sum_{\langle ij \rangle,
\alpha} i t (c_{\alpha,i}^\dagger c_{\alpha,j}-c_{\alpha,j}^\dagger c_{\alpha,i})\rangle_{\mathrm{EX}},
\label{jdef}
\end{equation}
where $\langle ij \rangle$ represents a sum over all links.  Note that the orientation and direction
of the link determines its contribution to the vector $\mathbf{j}$. We can interpret the excited state
trial wave function as the ground state for a system of $N$ particles to which we add one 
unpaired electron so that our state has a net spin and charge.  The Gutzwiller projection enforces 
strong correlation between the added electron and the $N$-electron ground state, so that we expect 
an effective quasiparticle with renormalized parameters.

\begin{figure}[!h]
\centerline{\includegraphics[width=2.5in]{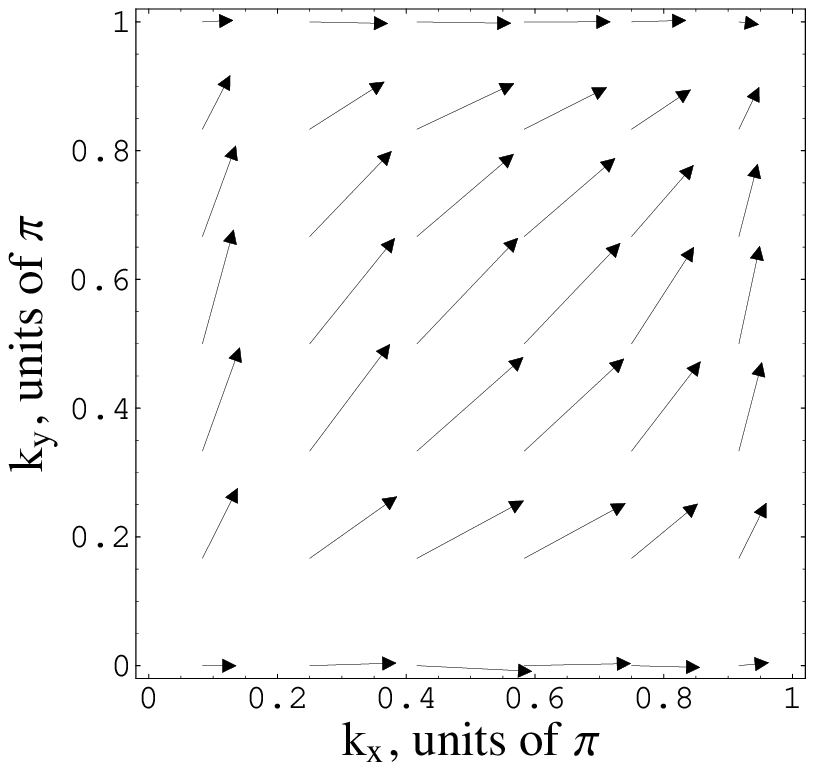}}
\centerline{\includegraphics[width=2.5in]{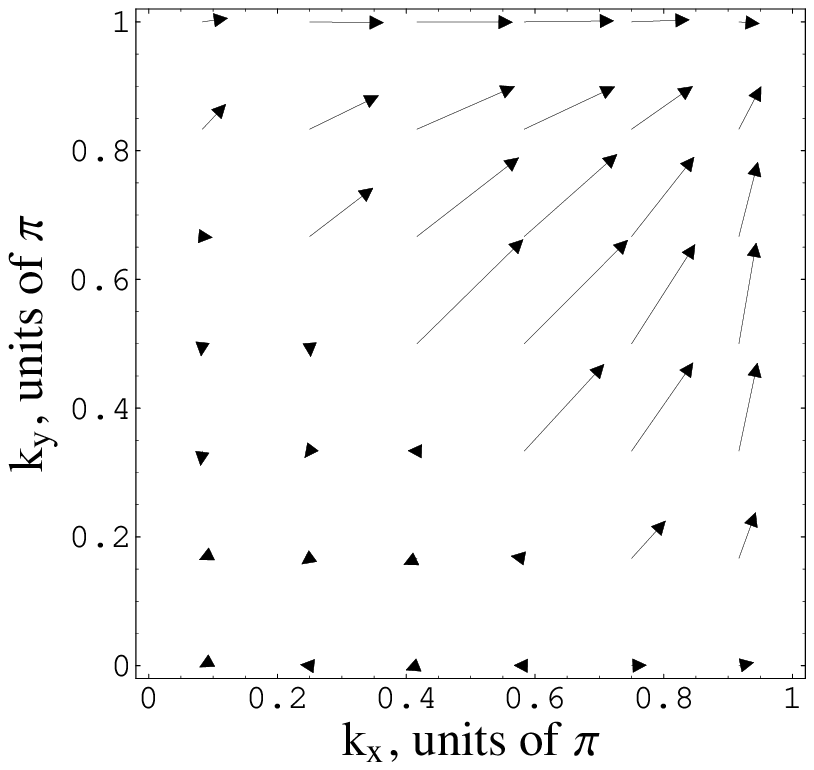}}
\centerline{\includegraphics[width=2.5in]{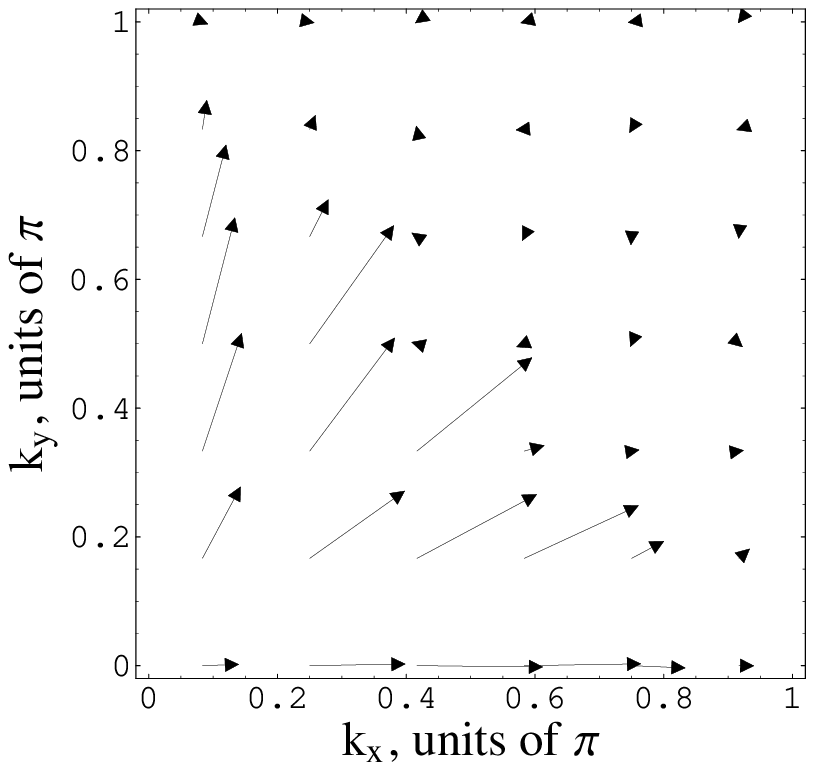}}
\caption{
The top middle and bottom figures are respectively plots of the current $\mathbf{j}$, $\mathbf{j_\uparrow}$, 
and $\mathbf{j_\downarrow}$ as a function of the wavevector $\mathbf{k}$ for a 12$\times$12 system with 13 
holes, $x=0.09$.  The vectors are drawn starting at the 
$\mathbf{k}$-point at which the current is calculated, their length is proportional to the
current magnitude. 
}
\label{jvec}
\end{figure}

\begin{figure}
\centerline{\includegraphics[width=3in]{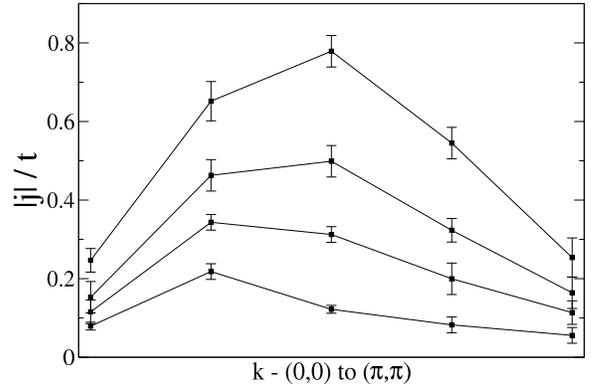}}
\caption{
The magnitude of the current $\left| \mathbf{j} \right|$ measured in units of $t$ along the nodal
diagonal, $(0,0)$ to $(\pi,\pi)$.  These runs were done on the 10$\times$10 system with the doping of 
0.01, 0.05, 0.09 and 0.17 (increasing magnitude).
}
\label{jnodaldiag}
\end{figure}

In the standard BCS theory (without the Gutzwiller projection), the current carried by the 
quasiparticles is $\mathbf{j}_{\mathbf{k}}=e\mathbf{v}_\mathbf{k}$ where
$\mathbf{v}_\mathbf{k}= d\varepsilon_\mathbf{k}/d\mathbf{k}$ 
is the velocity from the underlying normal state and not
the velocity from the quasiparticle dispersion $dE/d\mathbf{k}$.  This is the result of the fact that in BCS
theory the excitations are a superposition of particle and hole states and that these states carry
opposite charge but also move in opposite directions.  
The underlying metallic state can have some Fermi liquid correction to the current carried by the
particles and holes.  This correction is then carried through to the quasiparticle current in the
superconducting state, $\mathbf{j}_\mathbf{k}=\alpha e\mathbf{v}_\mathbf{k}$.  
In the case of the Gutzwiller projected
trial excited state that we are studying, we see that current does still approximately 
follow the shape of the
dispersion of the underlying metal and that the quasiparticle current renormalization $\alpha$ can be
calculated from the ratio of $\mathbf{j} / \mathbf{v}_\mathbf{k}$.

\subsection{Current as a function of k}
\label{subsection:current-k}

First we examine $\mathbf{j}$, the current carried by the quasiparticle, as a function of 
$\mathbf{k}$ for various dopings.  At most dopings,
we find a distribution with a structure similar to what one would expect from the 
tight binding model. 
In the top plot of Fig. \ref{jvec}, 
a typical example of the current carried by the quasiparticle as a function of wavevector is plotted. We
compare the direction of the current as a function of momentum to both the quasiparticle dispersion,
$dE(\mathbf{k})/d\mathbf{k}$, and to the underlying dispersion of the normal state $\mathbf{v}_\mathbf{k}$.
We find that the shape of the current indeed approximately follows the dispersion of the normal state
and not the dispersion of the quasiparticles.  
This is the same as in the BCS theory so the Gutzwiller
projection does not change this aspect of the physics.

For intermediate doping values, the magnitude of the current 
reaches its maximum value near the center of the  Brillouin zone, and
therefore near the nodal point. However, for strongly underdoped simulations, $x<0.05$, we find
that the maximum moves inward along the nodal direction, becoming closer to $(\pi/4,\pi/4)$ as
the doping approaches zero. (See Fig. \ref{jnodaldiag})

We also look at  how much of the total current is being carried by the up spins and down spins
individually.  By restricting the $\sum_{\alpha}$ in equation \ref{jdef} so that we consider 
$\alpha = \uparrow$ and $\alpha = \downarrow$ separately, we can investigate this property of
our trial excited state.  Since we are now distinguishing between these two spins, it is important 
to note
that we define our trial wave function as adding an up-spin to the system.
Given this, we find that the current of our state is almost entirely carried by either the up spins or 
the down spins depending on whether or not we are inside or outside of the effective Fermi surface as
can be seen in the lower two plots of Fig. \ref{jvec}.
Inside of the Fermi surface, all of the current is carried by the down spins and
outside of the Fermi surface all of the current is carried by the up spins.  
This is the qualitatively the same as in the unprojected case where the up and down spin currents have
factors of $u_\mathbf{k}^2$ and $v_\mathbf{k}^2$ respectively.

\begin{figure}[!ht]
\centerline{\includegraphics[width=2.5in]{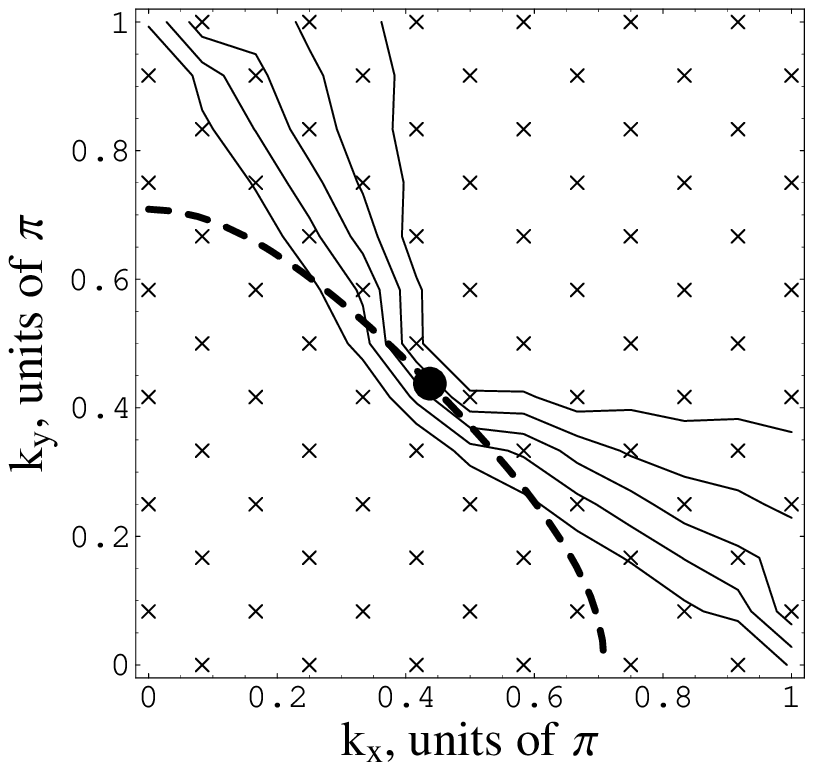}}
\centerline{\includegraphics[width=2.5in]{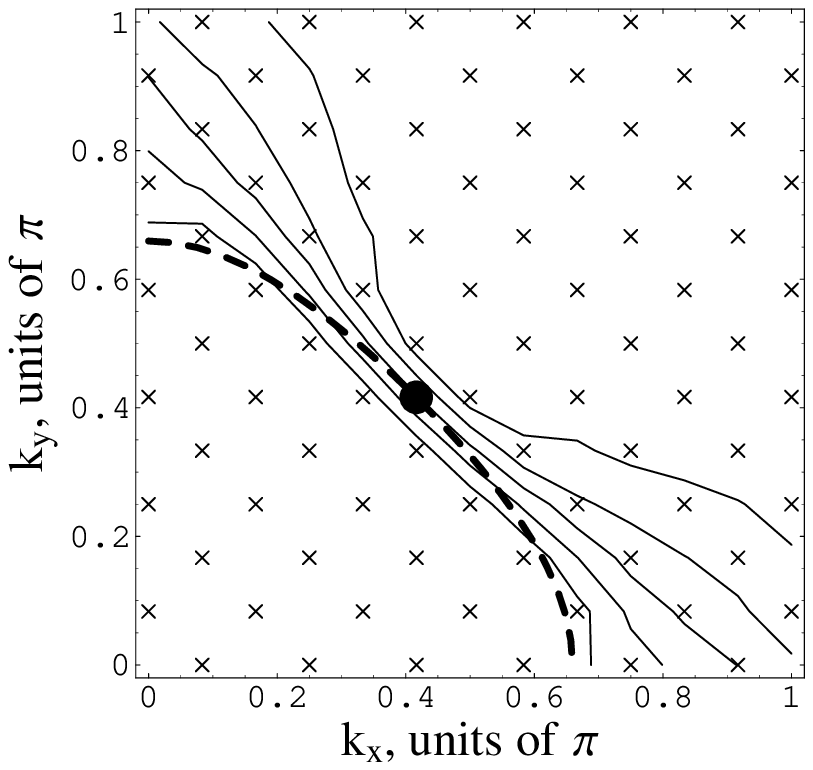}}
\caption{The particle contribution to the quasiparticle current
$n_\mathbf{j}(\mathbf{k})$ defined in Eq.\ (\protect\ref{nj-definition}).
The top and the bottom plots correspond to 13 and 31 holes in the 12$\times$12 system
(doping $x=0.09$ and $x=0.22$, respectively). The contour lines are
$n_\mathbf{j}(\mathbf{k})=0.1$, 0.3, 0.5, 0.7, and 0.9 (from left to right). The
thick dashed line is the Fermi surface of the unprojected state, the thick solid
dot marks the position of the node. Small crosses indicate the positions of data
points (we have used the reflection about the $x$--$y$ diagonal to double the
density of data points).
}
\label{fig:nj}
\end{figure}

\begin{figure}
\centerline{\includegraphics[width=3in]{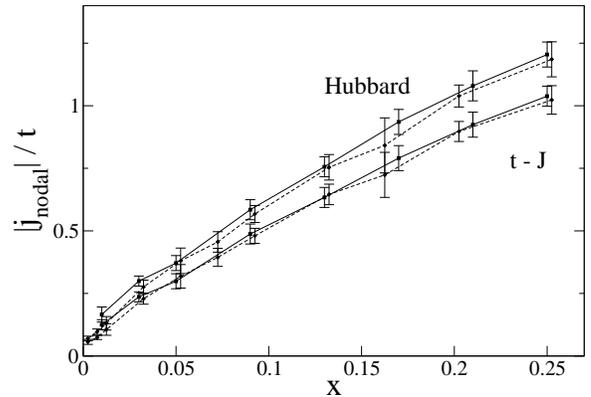}}
\caption{
Magnitude of the nodal current, $\left| \mathbf{j}_{\mathrm{nodal}} \right|$ measured in units
of $t$, plotted as a function of doping.
The solid lines are the data from a 10$\times$10 lattice and the dashed lines from a 20$\times$20 one. 
The lower curves are for the trial wavefunction for the $t$--$J$ model and the 
upper curves are for the trial 
wavefunction for the Hubbard model (to the lowest order in $t/U=1/12$).
}
\label{jvsx}
\end{figure}

To make a more detailed comparison to the BCS theory, we define the ``particle contribution''
to the current as
\begin{equation}
n_\mathbf{j} = \frac{\mathbf{j}_\downarrow(\mathbf{k}) \cdot 
\mathbf{j}_{\mathrm tot}(\mathbf{k})}{|\mathbf{j}_{\mathrm tot}(\mathbf{k})|^2} \, ,
\label{nj-definition}
\end{equation}
where $\mathbf{j}_{\mathrm tot} = \mathbf{j}_\downarrow + \mathbf{j}_\uparrow$ is the total current
carried by the quasiparticle. In the BCS theory, $n_\mathbf{j}(\mathbf{k}) = |v_\mathbf{k}|^2$,
it takes the values between zero and one, and the isoline $n_\mathbf{j}(\mathbf{k})=0.5$ coincides with
the Fermi surface. 
In Fig.\ \ref{fig:nj} we show the contour plots of  $n_\mathbf{j}(\mathbf{k})$ for
different values of doping, together with the Fermi surface for the corresponding
unprojected BCS wave functions. 
We see that the ``effective Fermi surface'' defined by  
$n_\mathbf{j}(\mathbf{k})$ does not follow the original Fermi surface of the BCS state,
but bends outwards in the $(0,\pi)$ regions.  Thus
it effectively acquires an inward curvature similar to the effect of the negative
$t'$ hopping term.

\subsection{Current as a function of doping}

We noted earlier that the magnitude of the current has a maximum near the nodal point and that
this maximum decreases as a function of doping. In Fig.\ \ref{jvsx}, we plot the magnitude of the nodal
current $\mathbf{j}_\mathrm{nodal}=\mathbf{j}(\mathbf{k}=(\pi/2,\pi/2))$ versus doping for 
both 10$\times$10 and 20$\times$20 systems.  Because of our choice of boundary conditions, 
we do not calculate the current at the true nodal point.
For a 10$\times$10 system, the ``nodal point'' is actually evaluated at $\mathbf{k}=(0.5\pi,0.4\pi)$ 
and for a  20$\times$20 system at $\mathbf{k}=(0.45\pi,0.5\pi)$.  
In Fig.\ \ref{jvsx}, we see that there is agreement to within the error between the data calculated for
two lattices of different size, and we expect that the actual nodal current would also be within
these errors.

In the 20$\times$20 system, we can study the doping as low as 0.005 (2 holes), and our 
results indicate that the current apparently decreases down to zero with decreasing
doping.

\subsection{Rotation to the Hubbard model}

So far we have studied the properties of fully projected wave functions.  
We would now like to extend our simulations from the $t$--$J$ model back
to the Hubbard model. This can be done in the standard way by employing
a unitary transformation $e^{iS}$ that decouples the Hilbert 
space of the Hubbard Hamiltonian so that there are no matrix 
elements connecting those subspaces with
different numbers of doubly occupied sites.  
Following MacDonald \textit{et al.}\cite{MacDonald:88}, we determine this
transformation as a power series in $(t/U)$, so that the 
subspaces are decoupled order by order.  To the first order, the rotation
generator is given by
\begin{equation}
iS = \frac{1}{U}\left(T_1 - T_{-1}\right) \, ,
\end{equation}
where $T_1$ and $T_{-1}$ are defined to be the parts of the kinetic 
energy operator that increase and decrease, respectively, 
the number of doubly occupied sites by one.

With the use of this rotation, we can replace computing the expectation
value of any operator $O$ in the ground state of a Hubbard model by
computing the expectation value of the rotated operator 
$\langle e^{i S} O e^{-i S} \rangle$ in the ground state of the
$t$--$J$ model. We use the same variational wave function for the $t$--$J$
model as described above and compute the lowest-order correction
(linear in $t/U$) to the $t$--$J$ expectation value.
This procedure has been applied, for example, in the work of
Paramekanti \textit{et al.} \cite{Paramekanti:04} for calculating the
expectation value of the occupation-number operator.

We compute the rotation correction to the value of the nodal current
and find that it does not qualitatively change its doping dependence.
The corrected value of the current is plotted as the upper
curves in Fig.\ \ref{jvsx} (for $t/U=1/12$) 

\subsection{Quasiparticle current renormalization $\alpha$}

Beyond just studying the current itself, we are particularly interested in the quasiparticle 
current renormalization factor $\alpha$. As we noted earlier the current and the slope of the
dispersion are collinear within the error of our simulations, so we can define 
$\alpha = \mathbf{j} / \mathbf{v}_\mathbf{k}$. 

We look at the quasiparticle current renormalization parameter $\alpha$ as a function of doping.  
We are interested in $\alpha$ for the lowest lying excitations, i.e., for those at the nodal point.  
Although we have calculated the Fermi velocity at the nodal point, more precise data for this
velocity as a function of doping are available from the work by Yunoki \textit{et al.} Ref. \cite{Sorella:05}.  
We use those Fermi velocity data in conjunction with our nodal current results to calculate $\alpha$ at the nodal
point.

\begin{figure}
\centerline{\includegraphics[width=3in]{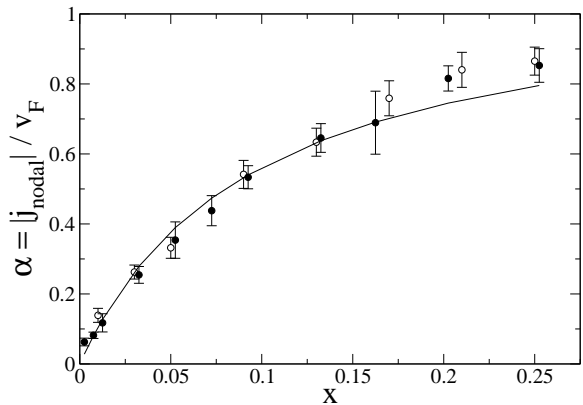}}
\caption{
The renormalization of the quasiparticle current $\alpha = |\mathbf{j}_{\mathrm{nodal}}|/v_F$ 
as a function of doping. The open circles are for runs on a 10$\times$10 lattice 
and the solid ones are for runs on a 20$\times$20 one. The solid line is the fit to the 20$\times$20 data by
Eq.\ (\ref{alpha-fit}) at $x^*=0.09$.
}
\label{fig:alphax}
\end{figure}

In Fig.~\ref{fig:alphax}, we plot the nodal value of $\alpha$ as a function of doping.  
We find that $\alpha$ goes to zero at zero doping.

It is interesting to compare our results with the predictions of slave-boson theory.  In this theory, 
the low lying excitations are $x$ bosons which carry charge with an effective hopping matrix element
proportional to $t$ and fermions which carry spin with an effecting hopping proportional to $J$.  These
excitations are coupled to gauge fluctuations.  When the gauge fluctuations are treated at the Gaussian
level, we obtain the Ioffe--Larkin composition rule which states that the inverse of the superfluid
density $\rho_s=n_s/m$ is given by adding the inverses of the fermion and boson contributions:
\begin{equation}
\rho_s^{-1}=(\rho_s^F)^{-1}+(\rho_s^B)^{-1}
\label{ILcomprule}
\end{equation}
where $\rho_s^B\approx xt$ and $\rho_s^F\approx J(1-aT)$ with $a\approx\Delta^{-1}$. \cite{Ioffe:89} 
Note that the linear temperature dependence comes from thermal excitations of the nodal fermions.
Expanding Eq. \ref{ILcomprule} at small $T$, we obtain
\begin{equation}
\rho_s(T) \approx  \rho_s(0) - \frac{[\rho_s(0)]^2}{\rho_s^F(0)}aT \, .
\label{pred1}
\end{equation}
On the other hand, the quasiparticle dispersion is given by that of the fermions and we can identify
$v_F/v_\Delta$ in Eq. \ref{maineq1} as being proportional to 
$aJ\sim a\rho_s^F(0)$.  
Comparison of Eq. \ref{pred1} with Eq. \ref{maineq1} results in the simple expression
\begin{equation}
\alpha\sim\frac{\rho_s(0)}{\rho_s^F(0)}\, ,
\label{pred2}
\end{equation}
in particular $\alpha\propto x(t/J)$ for $xt<J$. \cite{Lee:05}  
We did not keep track of the numerical coefficients. If we assume that at full doping ($x=1$)
$\alpha$ should approach one (the BCS value), Eq.\ (\ref{pred2}) suggests the form
\begin{equation}
\alpha(x)=\frac{x}{x+x^*} (1+x^*)\, ,
\label{alpha-fit}
\end{equation}
which for $x^*=0.09$ produces a qualitatively good fit of our results for $\alpha$ (see Fig.~\ref{fig:alphax}).
Note that the order-of-magnitude estimates above gives $x^* \sim J/t$, and our
best-fit value of $x^*$ is several times smaller.

The shape and the magnitude of the current renormalization parameter $\alpha$
as a function of doping is currently a topic of
much experimental work.  While there still remain large uncertainties in the experimental data,
two useful comparisons can be made to our work.  In Ref. \onlinecite{Chiao:00},
the measurements of thermal conductivity and of the penetration depth were used to determine
the ratio $v_F/v_\Delta$ and the superfluid density. Combining those data resulted in the
the values of $\alpha=0.66$ and $\alpha=0.68$ for optimally doped samples of BSCCO
and YBCO, respectively. Those numbers are in close agreement with our results around $x=0.15$ doping.  

Our results also qualitatively agree with the decrease of $\alpha$ as the doping decreases in
underdoped YBCO, as reported in Refs.\ \onlinecite{Liang:05,Broun:05}.
However, earlier data on YBCO films indicated that the linear $T$ slope of $n_s/m$ is relatively
insensitive to doping over a broad range of critical temperatures. \cite{Boyce:00,Stajic:03}
Since $v_F/v_\Delta$ decreases with decreasing $x$ \cite{Sutherland:03}, this trend 
is in disagreement with Fig.~\ref{fig:alphax}
and Eq.\ (\ref{alpha-fit}).  It will be desirable to have single crystal data for $n_s(T)/m$ over a 
broad range of $x$ to settle this point.

On the theoretical side, the recent study of a $U(1)$ slave-boson theory with spinon-holon binding \cite{Ng:05}
predicted a sub-linear dependence of $\alpha$ on the doping. The sublinear form of $\alpha(x)$ disagrees
with our proposal (\ref{alpha-fit}), but is also consistent with our numerical results shown
in Fig.~\ref{fig:alphax}.

\begin{figure}
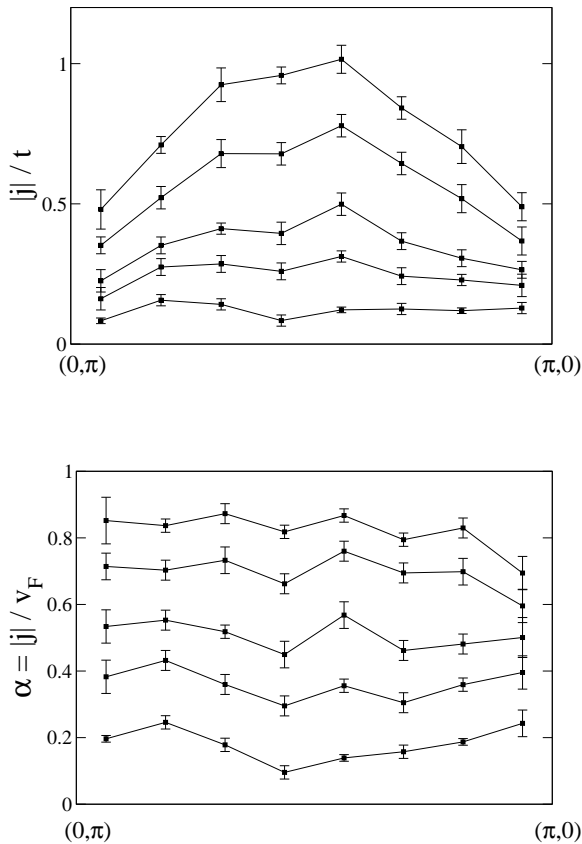

\centerline{\includegraphics[width=3in]{jfsfinal2.eps}}
$\_$\newline
\newline
\newline
\centerline{\includegraphics[width=3in]{afsfinal2.eps}}
\caption{
Plot of $\left| \mathbf{j} \right|$ and $\alpha = \left| \mathbf{j} \right| / v_F$, upper
and lower plots respectively, for the $t$--$J$ model along the Fermi surface for dopings of 
0.01,0.05,0.09,0.17 and 0.25 (increasing magnitude).  
}
\label{jandafs}
\end{figure}

We further look at $\alpha$ along the Fermi surface for a given doping.  We are interested
in how the shape and size of this curve changes as a function of doping.  While we expect the integral of the
whole curve to decrease with $x$, there are several possibilities for how this could occur.  
Two such scenarios are
that the curve decreases in magnitude everywhere along the Fermi surface uniformly as doping decreases or
that $\alpha$ is small almost everywhere along the Fermi surface but that there is a region of large
$\alpha$ around the nodal point whose width
increases with $x$.  Our numerical results indicate the former of those scenarios.
In the upper graph of Fig. \ref{jandafs}, we plot the 
$t$--$J$ model current magnitude along the Fermi surface.  Due to the low finite resolution of our system,
we just calculate the current at those points nearest to the line connecting $(\pi,0)$ and $(0,\pi)$.  
To calculate $\alpha$, we use the value of $v_F$ found by Yunoki \textit{et al.} \cite{Sorella:05} 
along the nodal direction for a given doping and accordingly renormalize the tight binding dispersion to 
calculate $\mathbf{v}_\mathbf{k}$ along these points.  We plot $\alpha$ obtained in this way in 
the lower plot of Fig. \ref{jandafs}. We see that $\alpha$ is approximately flat along the Fermi surface.

\section{Quasiparticle weight and Occupation Number}

\begin{figure}
\centerline{\includegraphics[width=2.5in]{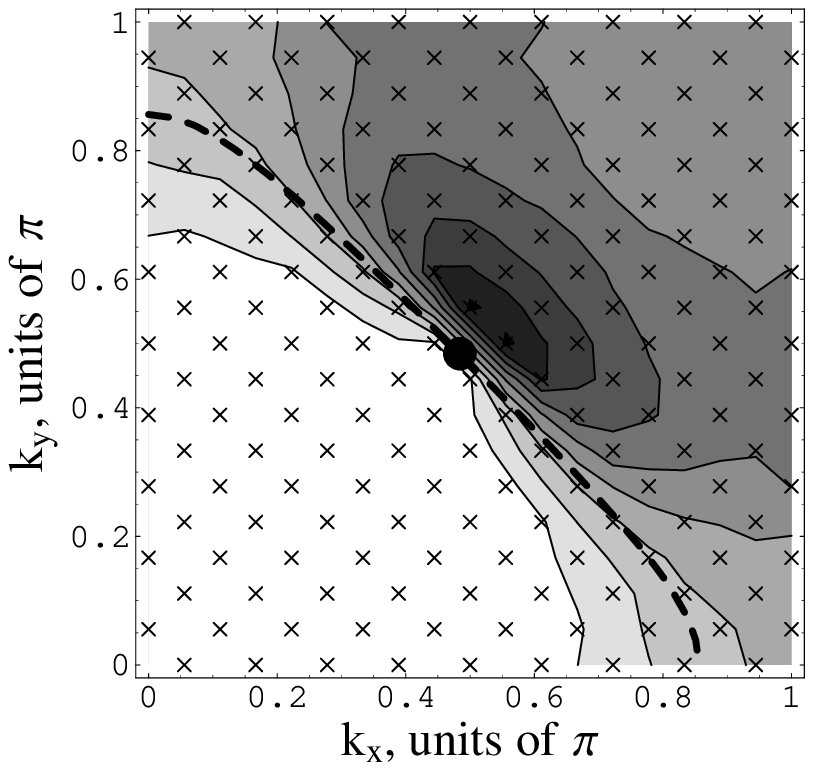}}
\centerline{\includegraphics[width=2.5in]{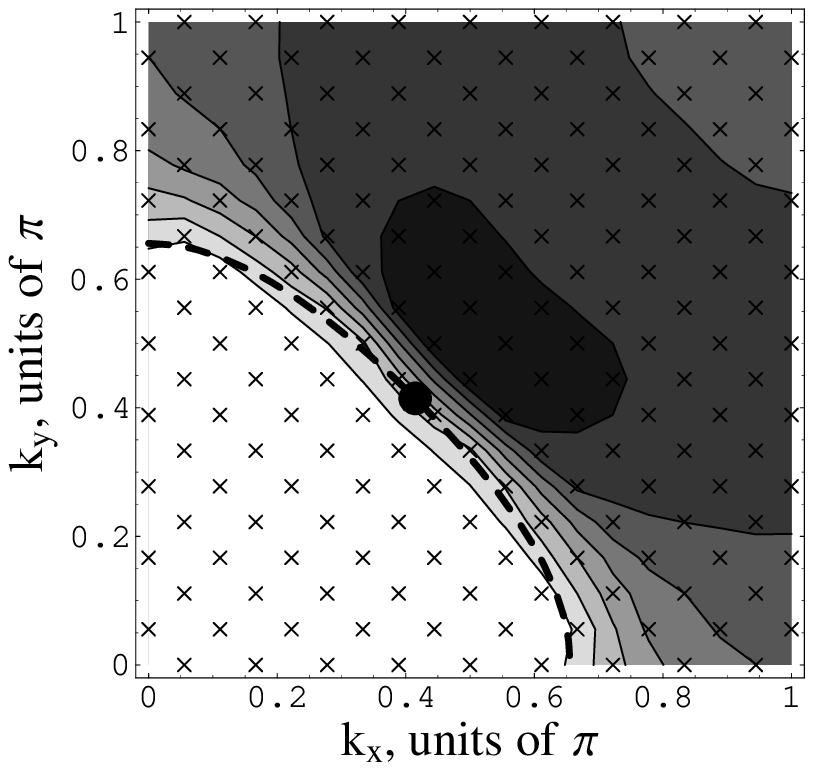}}
\caption{
Contour plots of $Z_+(\mathbf{k})$ in the $t$--$J$ model
for dopings 0.03 (upper plot) and 0.21 (lower plot),
10 and 68 holes in a 18$\times$18 system, respectively.
The thick dashed lines denote the Fermi surface of the unprojected
wave function, the big solid dot marks the position of the node.
Crosses indicate data points used in plotting (available values of the
$\mathbf{k}$ vector). In the upper plot, the contour lines correspond to
$Z_+=0.01$, $0.02$, $\dots$, $0.09$ (left to right -- the maximal value of $Z_+$ is $0.09$).
In the lower plot, the contour lines are $Z_+=0.05$, $0.10$, $\dots$,
$0.35$ (with the maximal value $Z_+=0.38$).
}
\label{fig:Z+}
\end{figure}

In this section we examine the quasiparticle weight $Z$ and the occupation number $n$ as a 
function of momentum and doping.  The quasiparticle weight is defined as in Fermi liquid 
theory and gives a measure of how close our trial wave function quasiparticle is to being a free
electron (or a free hole).  

We begin with the definitions,
\begin{eqnarray}
\label{zpluseq}
&&Z_+(\mathbf{k},\sigma) =\frac{\left |\langle\Psi_{\mathrm{EX}}|c_{\mathbf{k},\sigma}^{\dagger}
|\Psi_{\mathrm{GS}}\rangle \right|^2}{\langle\Psi_{\mathrm{EX}}|\Psi_{\mathrm{EX}}\rangle\langle
\Psi_{\mathrm{GS}}|\Psi_{\mathrm{GS}}\rangle}\, ,
\label{Z+def}\\
&&Z_-(\mathbf{k},\sigma) =\frac{\left |\langle\Psi_{\mathrm{EX}} |c_{\mathbf{k},\sigma}
|\Psi_{\mathrm{GS}}\rangle \right|^2}{\langle\Psi_{\mathrm{EX}}|\Psi_{\mathrm{EX}}\rangle\langle
\Psi_{\mathrm{GS}}|\Psi_{\mathrm{GS}}\rangle}\, ,
\label{Z-def}
\end{eqnarray}
where $|\Psi_{\mathrm{EX}}\rangle$ also carries momentum $\mathbf{k}$, and the
electron operators are normalized as $\{ c_{\mathbf{k},\sigma}, c_{\mathbf{k},\sigma}^{\dagger}\}=1$.
In the BCS theory (without Gutzwiller projection),
$Z_+=u_\mathbf{k}^2$ and $Z_-=v_\mathbf{k}^2$.  Along the nodal diagonal (where the gap vanishes), 
$u_\mathbf{k}^2=1$ outside 
the Fermi surface and $0$ inside and $v_\mathbf{k}^2$ is the opposite.  
If one defines $Z=Z_+ +Z_-$, then in the BCS model $Z=1$ everywhere. 

\begin{figure}
\centerline{\includegraphics[width=2.5in]{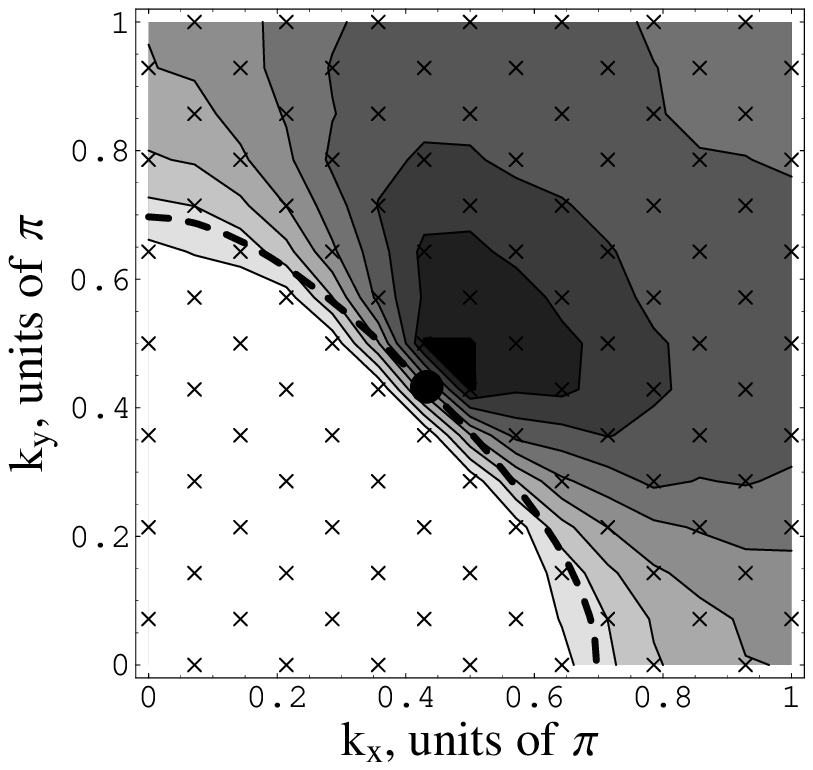}}
\centerline{\includegraphics[width=2.5in]{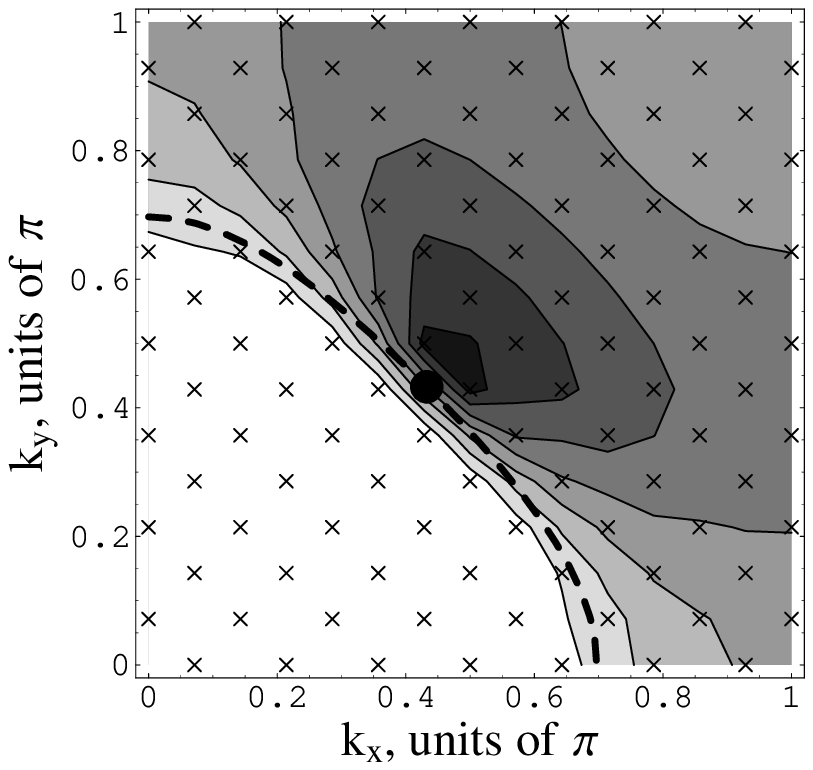}}
\centerline{\includegraphics[width=2.5in]{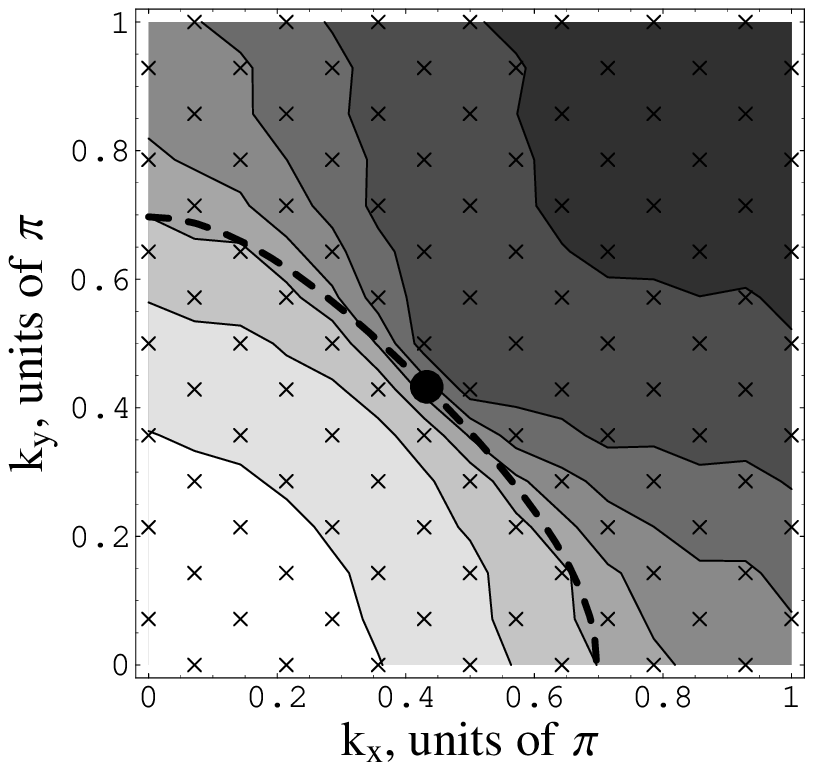}}
\caption{
{\bf Upper plot}: contour plot of $Z_+(\mathbf{k})$ (or, equivalently,
of $n(\mathbf{k})$) in the $t$--$J$ model
at the doping 0.12 (24 holes in the 14$\times$14 system).
The isolines correspond to $Z_+ = 0.03, 0.06,\dots, 0.27$ 
[$n(\mathbf{k})= 0.53, 0.50,\dots, 0.29$ respectively, left to right]. The maximal
value of $Z_+$ (near the node) is $0.27$.
{\bf Middle plot}: $Z_+^H$ in the same system rotated to the Hubbard model (to the
first order in $t/U=1/12$), Eq.\ (\protect\ref{rotzplus}). The isolines correspond to 
$Z_+^H = 0.05, 0.10,\dots, 0.35$. The maximal
value of $Z_+^H$ (near the node) is $0.37$.
{\bf Bottom plot}: $n_\mathbf{k}^H$ in the same system rotated to the
Hubbard model (to the first order in $t/U=1/12$), Eq.\ (\protect\ref{rotnk}). 
The isolines are  $n_\mathbf{k}^H = 0.8, 0.7,\dots, 0.2$ (left to right).
The Fermi surface, the nodal point and the positions of data points
are denoted in the same way as in Fig.\ \protect\ref{fig:Z+}.
}
\label{fig:Zrot}
\end{figure}

In the projected wave functions, these simple expressions do not apply. However, some of the
properties of the spectral weights $Z_+$ and $Z_-$ may be still proven. Using the identity
\begin{equation}
P_D c_k^{\dagger} P_D = P_D c_k^{\dagger}
\end{equation}
and the $d$-wave symmetry of the gap, we can prove that along the nodal diagonal,
\begin{eqnarray}
Z_+=0 &{}\quad {\rm on}\quad{}& (0,0) - (k_F,k_F)\, , 
\label{Z+zero}\\
Z_-=0 &{}\quad {\rm on}\quad{}& (k_F,k_F) - (\pi,\pi)\, ,
\label{Z-zero}
\end{eqnarray}
where $(k_F,k_F)$ is the nodal point. Furthermore, $Z_+$ may be rewritten as the ground-state
expectation value
\begin{equation}
Z_+(\mathbf{k},\sigma)=\langle c_{\mathbf{k},\sigma}P_Dc_{\mathbf{k},
\sigma}^{\dagger}\rangle_{\mathrm{GS}}.
\label{zdef}
\end{equation}
We note that
\begin{eqnarray}
\langle c_{i,\sigma}c_{j,\sigma}^{\dagger}\rangle_{\mathrm{GS}} &=& 
\langle c_{i,\sigma}Pc_{j,\sigma}^{\dagger}\rangle_{\mathrm{GS}}  \quad i\not=j\\
&=& \langle c_{i,\sigma}Pc_{j,\sigma}^{\dagger}\rangle_{\mathrm{GS}} + n_{\overline{\sigma}} \quad i=j.
\end{eqnarray}
Thus $Z_+({\mathbf{k},\sigma})$ is further related to the occupation number
\begin{equation}
n_{\mathbf{k},\sigma} = \langle c_{\mathbf{k},\sigma}^{\dagger} c_{\mathbf{k},\sigma} \rangle
\label{ndef}
\end{equation}
as
\begin{equation}
Z_+(\mathbf{k},\sigma)=\frac{1+x}{2}-n_{\mathbf{k},\sigma}\, .
\label{nzrelationeq}
\end{equation}
This relation has also been given by Yunoki in Ref.\ \onlinecite{Yunoki:05}.
In particular, from this relation follows the upper bound on the spectral weight $Z_+$:
\begin{equation}
Z_+\le \frac{1+x}{2}
\label{zbound}
\end{equation}
and, as a consequence, the same upper bound applies to the nodal spectral weight 
$Z_{\rm nodal}=Z_+(k_F+\epsilon,k_F+\epsilon)$ studied by Paramekanti \textit{et al.} in 
Refs.\ \onlinecite{Paramekanti:01,Paramekanti:04}.

The quasiparticle weight $Z_-$ requires a more complicated Monte Carlo calculation,
since it cannot be rewritten as a simple ground-state average like (\ref{zdef}).
We therefore restrict ourselves to discussing only the quasiparticle weight $Z_+$ in this paper.

The above relations for $Z_+$ have been derived for the fully projected wave function.
If we perform a rotation to the Hubbard model, the relations (\ref{zdef}) and (\ref{nzrelationeq})
no longer hold, and the upper bound (\ref{zbound}) cannot be proven. Specifically, for the
Hubbard model we keep the same definitions of the quasiparticle weights and the
occupation number (\ref{Z+def}), (\ref{Z-def}), (\ref{ndef}), but with the ground and
excited states rotated from the fully-projected state to the Hubbard-model state
by the unitary rotation $e^{-iS}$, as explained in the previous section.
Then the lowest-order Hubbard-model corrections to $Z_+$ and to $n_{\mathbf{k}}$ may be easily
computed as 
\begin{equation}
Z_+^H = Z_+ + 2\; {\rm Re}\; 
\langle c_{\mathbf{k},\sigma} P_D \lbrack iS,c^\dagger_{\mathbf{k},\sigma} \rbrack \rangle
\label{rotzplus}
\end{equation}
and
\begin{equation}
n_{\mathbf{k}}^H= n_{\mathbf{k}} - 2\; {\rm Re}\; 
\langle c_{\mathbf{k},\sigma} \lbrack iS,c^\dagger_{\mathbf{k},\sigma} \rbrack \rangle\, .
\label{rotnk}
\end{equation}
Even though the two expressions look nearly identical, the correction to $n_{\mathbf{k}}$ does
not contain an intermediate projector $P_D$ and, because of that, has a very different structure than
that to $Z_+$.  The relation between $Z_+$ and $n_\mathbf{k}$ (\ref{nzrelationeq}) no longer holds for 
the Hubbard model, as we shall see below.

Figs.\ \ref{fig:Z+} and \ref{fig:Zrot} show $Z_+(\mathbf{k})$ for 
$x=0.03$, $0.12$, and $0.21$. As noted in Eq.\ (\ref{Z+zero}),
$Z_+$ is zero along the diagonal between $(0,0)$ and $(k_F,k_F)$ and jumps to a finite
value $Z_\mathrm{nodal}$ at the nodal point. The projected wave function inherits the essential singularity
of $Z_+$ at the nodal point from the underlying BCS wave function. The value of $Z_\mathrm{nodal}$
has been studied as a function of doping by Paramekanti \textit{et al.} in 
Refs.\ \onlinecite{Paramekanti:01,Paramekanti:04}. The doping dependence of $Z_\mathrm{nodal}$ 
(Fig.~2 of Ref.\ \onlinecite{Paramekanti:01} and Fig.~6 of Ref.\ \onlinecite{Paramekanti:04})
is qualitatively similar to that we find for $\alpha$ (Fig.\ \ref{fig:alphax}): 
both $Z_\mathrm{nodal}$ and $\alpha$ decrease to zero with decreasing doping, with a strong upward curvature.
However we are not aware of any a priori relation between $\alpha$ and $Z_\mathrm{nodal}$.

Using (\ref{nzrelationeq}), the plots of $Z_+(\mathbf{k})$ may also be interpreted as those of $n_\mathbf{k}$
(see, e.g., the upper plot in Fig.\ \ref{fig:Zrot}).  Note the region of 
depression in $n_\mathbf{k}$ just outside the Fermi surface, which resembles a hole ``pocket.''
The existence of this pocket may already be inferred from the non-monotonous behavior of $n_\mathbf{k}$
along the zone diagonal found in Refs.\ \onlinecite{Paramekanti:01,Paramekanti:04} but
the full $\mathbf{k}$ dependence shown in Figs.\ \ref{fig:Z+} and \ref{fig:Zrot} gives a more complete picture.
Remarkably, a similar ``pocket'' structure has been found in the $U(1)$ slave-boson model with
spinon-holon-binding by Ng \cite{Ng:05}. The pocket is more pronounced at lower dopings and
appears consistent with the bending of the ``effective Fermi surface'' defined in Section
\ref{section:current} from $n_\mathbf{j}$, Eq.\ \ref{nj-definition}. 
However, to define a meaningful Fermi surface from the quasiparticle
spectral weight, one needs an access also to the spectral weight $Z_-$ which goes beyond
the scope of the present paper.

\begin{figure}
\centerline{\includegraphics[width=3in]{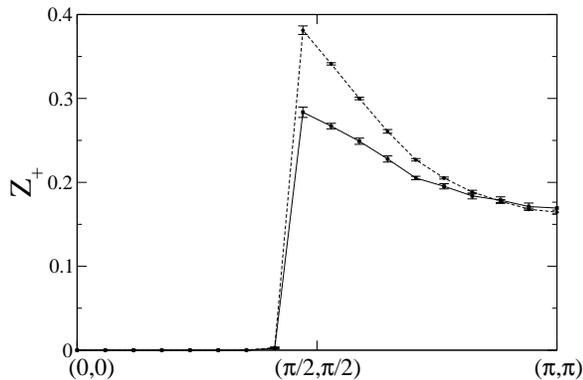}}
\caption{Plot of $Z_+$ and $Z_+^H$, the solid and dashed lines respectively, 
for an 18$\times$18 system with 40 holes, $x=0.12$.  The data are calculated
on the discrete grid points closest to the diagonal.
}
\label{fig:Znodal}
\end{figure}

If we include the $t/U$ correction from the rotation to the Hubbard model, the
pocket structure in $n_\mathbf{k}^H$ disappears, see our Fig.\ \ref{fig:Zrot} (lower plot) and
Refs.\ \onlinecite{Paramekanti:01,Paramekanti:04}. On the other hand, including the $t/U$
correction to $Z_+({\mathbf{k}})$ preserves the pocket structure, see Fig.\ \ref{fig:Zrot} (middle plot).
This shows that in the Hubbard model the relation (\ref{nzrelationeq})
between $Z_+^H({\mathbf{k}})$ and $n_\mathbf{k}^H$ no longer holds.

In Refs.\ \onlinecite{Paramekanti:01,Paramekanti:04}, it was reported that the rotation to the 
Hubbard model does not change the magnitude of the the jump in $n_\mathbf{k}$ at the Fermi surface.
Our results on $n_\mathbf{k}$ and $n_\mathbf{k}^H$ confirm this statement, however the
spectral weight defined as $Z_+$ increases when rotated to the Hubbard model.
In Fig.\ \ref{fig:Znodal}, we show $Z_+$ and $Z_+^H$ along the nodal diagonal for an $x=0.12$
system.

\section{Conclusion}
In this paper, we have analyzed the properties of the excited states in the $t$--$J$ and
Hubbard models in the framework of Gutzwiller-projected variational wave functions. 
The quantities of main interest are the renormalization of the current and spectral
weight of the quasiparticles. Both those renormalizations decrease with decreasing
doping and exhibit strongly non-BCS behavior. 

The renormalization of the quasiparticle
current allows us to define the effective Fermi surface as a crossover region between
the electron- and hole-supported current. We observe that such a Fermi surface bends
outwards in the $(0,\pi)$ regions -- an effect normally ascribed to the $t'$ hopping term.
At the same time, the total current is renormalized approximately uniformly along the
Fermi surface. 

The renormalization of the quasiparticle spectral weight, on the other hand, is peaked
near the nodal point and exhibits a pocket-like structure. 
This pocket feature is more pronounced at lower doping values.

Comparing the results for the $t$--$J$ model and for the Hubbard model (to the lowest
order in the $t/U$ correction), we find that the rotation to the Hubbard model does not
qualitatively affect the renormalizations of the quasiparticle spectral
weight and of the current.

We would like to thank Professor T.K. Lee and Professor Castellani for their discussions and help
on this research. C.P.N.\ and P.A.L.\ acknowledge support by NSF grant DMR-0517222. 

\bibliography{tjbib}

\end{document}